\documentclass[prd,twocolumn,showpacs,preprintnumbers,amsmath,amssymb]{revtex4}

\usepackage{amsmath}
\usepackage{graphicx}
\usepackage{dcolumn}
\usepackage{xspace}
\usepackage{color}
\usepackage{url}
\usepackage[utf8]{inputenc}
\usepackage{epstopdf}

\allowdisplaybreaks

\newcommand{\bea}{\begin{eqnarray}}
\newcommand{\eea}{\end{eqnarray}}
\newcommand{\nn}{\nonumber\\}

\begin{document}
%%%%%%%%%%%%%%%%%%%%%%%%%%%%%%%%%
\preprint{PSI-PR-18-01, TTP18-005}
\title{\boldmath $B$ Meson Anomalies in a Pati-Salam Model within the Randall-Sundrum Background}

\author{Monika Blanke}
\email{monika.blanke@kit.edu}
\affiliation{Institut f\"ur Kernphysik, Karlsruhe Institute of Technology,
  Hermann-von-Helmholtz-Platz 1,
  D-76344 Eggenstein-Leopoldshafen, Germany\vspace{.3mm}\\
 Institut f\"ur Theoretische Teilchenphysik,
  Karlsruhe Institute of Technology, Engesserstra\ss e 7,
  D-76128 Karlsruhe, Germany}

\author{Andreas Crivellin}
\email{andreas.crivellin@cern.ch}
\affiliation{Paul Scherrer Institut, CH--5232 Villigen PSI, Switzerland}

%%%%%%%%%%%%%%%%%%%%%%%%%%%%%%%%%%
%%%%%%%%%%%%%%%%%%%%%%%%%%%%%%%%%%
\begin{abstract}
Lepton number as a fourth color is the intriguing theoretical idea of the famous Pati-Salam (PS) model. While in conventional PS models, the symmetry breaking scale and the mass of the resulting vector leptoquark are stringently constrained by $K_L\to\mu e$ and $K\to\pi\mu e$, the scale can be lowered to a few TeV by {adding vector-like fermions. Furthermore, in this case, the intriguing hints for lepton flavour universality violation in $b\to s\mu^+\mu^-$ and $b\to c\tau\nu$ processes can be addressed. Such a setup is naturally achieved by implementing the PS gauge group in the five-dimensional Randall-Sundrum background. The PS symmetry is broken by boundary conditions on the fifth dimension and the resulting massive vector leptoquark automatically has  the same mass scale as the vector-like fermions and all other resonances. We consider the phenomenology of this model in the context of the hints for lepton flavour universality violation in semileptonic $B$ decays. Assuming flavour alignment in the down sector we find that in $b\to s\ell^+\ell^-$ transitions the observed deviations from the SM predictions (including $R(K)$ and $R(K^*)$) can be explained with natural values for the free parameters of the model.} Even though we find sizable effects in $R(D)$, $R(D^*)$ and $R(J/\Psi)$ one cannot account for the current central values in the constrained setup of our minimal model due to the stringent constraints from $D-\bar D$ mixing and $\tau\to 3\mu$. 
\end{abstract}

%%%%%%%%%%%%%%%%%%%%%%%%%%%%%%%%%
%%%%%%%%%%%%%%%%%%%%%%%%%%%%%%%%%
\maketitle

%%%%%%%%%%%
\section{Introduction}
\label{intro}
%%%%%%%%%%%

So far, the Large Hadron Collider (LHC) at CERN did not directly observe any particles beyond the ones of the {Standard Model (SM) of particle physics. However, we have accumulated intriguing hints for lepton flavor universality (LFU) violation in semi-leptonic $B$ decays within recent years. Most prominently, there exist deviations from the SM predictions in $b\to s\mu^+\mu^-$ above the $5\,\sigma$ level \cite{Capdevila:2017bsm} \footnote{Including only $R(K)$ and $R(K^*)$ the significance is at the 4$\,\sigma$ level~\cite{Altmannshofer:2017yso,DAmico:2017mtc,Geng:2017svp,Ciuchini:2017mik,Hiller:2017bzc,Hurth:2017hxg}.} and the combination of the ratios $R(D)$ and $R(D^*)$ differs by $4.1\,\sigma$ from its SM prediction~\cite{Amhis:2016xyh}. Furthermore, also $R(J/\Psi)$ points towards the violation of LFU in $b\to c\tau\nu$  processes \cite{Aaij:2017tyk}. This suggests a possible connection between these two classes of decays and motivates the investigation of simultaneous explanations ~\cite{Bhattacharya:2014wla,Alonso:2015sja,Calibbi:2015kma,Fajfer:2015ycq,Greljo:2015mma,Barbieri:2015yvd,Bauer:2015knc,Boucenna:2016qad,Das:2016vkr,Becirevic:2016yqi,Sahoo:2016pet,Bhattacharya:2016mcc,Barbieri:2016las,Crivellin:2017zlb,Chen:2017hir,Dorsner:2017ufx,Buttazzo:2017ixm,DiLuzio:2017vat,Bordone:2017bld,Barbieri:2017tuq}. 

In fact, the $SU(2)$ singlet vector leptoquark (VLQ) with hypercharge $2/3$ is a natural candidate for a simultaneous explanation \cite{Alonso:2015sja,Calibbi:2015kma,Barbieri:2015yvd,Buttazzo:2017ixm}. It contributes to $b\to s\mu^+\mu^-$ as well as to $b\to c\tau\nu$ and it does not couple down-quarks to neutrinos, avoiding the bounds from $B\to K^{(*)}\nu\nu$ and is free of proton decay to all orders in perturbation theory~\cite{Assad:2017iib}. This allows for large flavour violating effects and the bounds from direct searches~\cite{Faroughy:2016osc} and EW precision data~\cite{Feruglio:2016gvd,Feruglio:2017rjo} can be avoided~\cite{Crivellin:2017zlb,Buttazzo:2017ixm}. Interestingly, this LQ appears in the theoretically very appealing  Pati-Salam (PS)~\cite{Pati:1974yy} model and several attempts have been made in the literature to construct a model {addressing the flavour} anomalies based on the corresponding gauge symmetry~\cite{DiLuzio:2017vat,Calibbi:2017qbu,Bordone:2017bld,Barbieri:2017tuq}.

In conventional PS models, the bounds on the breaking scale from $K_L\to\mu e$ and $K\to\pi\mu e$ are so strong (at the PeV scale)~\cite{Hung:1981pd,Valencia:1994cj} that any other observable effect in flavour physics is ruled out. Nonetheless, if the PS gauge symmetry is implemented in the 5D Randall-Sundrum (RS) background \cite{Randall:1999ee}, the mass scale of the Kaluza-Klein (KK) resonances (including the VLQ) can be much lower, i.e. in the few TeV range \footnote{The $B$ physics anomalies have been considered in the context of ordinary composite Higgs or extra dimension models in Refs.~\cite{Niehoff:2015bfa,Carmona:2015ena,Megias:2016bde,Megias:2017ove,Carmona:2017fsn,Sannino:2017utc,DAmbrosio:2017wis}. For general flavor analyses of conventional RS models, see Refs.~\cite{Agashe:2004cp,Csaki:2008zd,Blanke:2008zb,Blanke:2008yr,Bauer:2009cf}.}. The suppression of the lepton flavour violating kaon decays can be achieved by introducing the SM fermions as zero modes of bulk fermions~\cite{Grossman:1999ra} with their couplings to the KK modes determined by their localization along the RS bulk. Since the zero mode localizations are free parameters, one can obtain the required non-trivial flavour structure in order to give interesting effects in $b\to s\mu^+\mu^-$ and $b\to c\tau\nu$ transitions. 

\section{The Model}

Our starting point is a 5D RS space-time~\cite{Randall:1999ee} 
\begin{equation}
ds^2 = e^{-2ky}\eta_{\mu\nu}dx^\mu dx^\nu - dy^2\,, \qquad 0\le y\le \pi R
\end{equation}
with the PS~\cite{Pati:1974yy} bulk gauge symmetry $SU(4)\times SU(2)_L\times SU(2)_R$. The symmetry is broken to its SM subgroup by means of boundary conditions on the UV brane. Note that the unbroken $U(1)_Y$ is a linear combination of $U(1)_{B-L}$ (contained in $SU(4)$) and $U(1)_R$ of $SU(2)_R$. Therefore, relaxing the assumption of a discrete left-right symmetry, one can always account for the measured values of $g_Y$

As in the conventional PS model, the SM fermions are embedded into complete representations of the PS gauge group. In addition, they are introduced as bulk fields in the RS background and the zero modes correspond to the SM quarks and leptons. Their localizations are determined by their 5D bulk masses \cite{Grossman:1999ra}, but can be altered by the presence of brane-kinetic terms. Since on the UV brane only the SM gauge symmetry is unbroken, the localizations of quarks and leptons of the same generation can differ from each other \footnote{We thank Csaba Cs\'aki for reminding us of this possibility.}. The Higgs doublet is introduced as a 4D field confined to the UV brane, hence its couplings to the KK modes are strongly suppressed. This choice ensures the compliance with electroweak precision constraints. 

The 4D dual theory, according to the AdS/CFT correspondence~\cite{Maldacena:1997re}, is a composite model with a global PS symmetry and KK modes corresponding to composite resonances. The gauging of the SM subgroup explicitly breaks the global symmetry in the elementary sector. Hence, the SM fermions are partially composite {due to} a linear mixing of the elementary fermions with composite operators of the same quantum numbers.
Therefore, the simplified version of our model, according to the deconstruction approach~\cite{Contino:2006nn},  contains composite vector resonances of the $SU(4)\times SU(2)_L\times SU(2)_R$ symmetry group with common mass $M$, as well as three generations of heavy vector-like quarks and leptons corresponding to the first KK modes.

%\subsection{Fermion masses}

Concerning fermions we have three generations of chiral (SM) fermions, the quark doublets $q_i$, the lepton doublets $\ell_i$, the quark singlets $d_i$ and $u_i$ as well as the lepton singlet $e_i$ (and the right-handed neutrino which we do not consider in the following as it is not relevant for our discussion). In addition, we have the three generations of vector-like fermions which we denote by the corresponding capital letters. The mass terms for the fermion before electroweak symmetry breaking read
\begin{eqnarray}
{\cal L}_M &=& -M_{ij}^L\left( \bar Q_i^L Q_j^R + \bar L_i^LL_j^R\right)\nonumber\\
&&-M_{ij}^R\left(\bar U_i^LU_j^R +\bar D_i^LD_j^R + \bar E_i^LE_j^R\right)\nonumber\\
 && - m_{ij}^{qL}{{\bar q}_i}Q_j^R - m_{ij}^{\ell L} {{\bar \ell }_i}L_j^R  \\
&& - m_{ij}^{uR} \bar U_i^L{u_j} - m_{ij}^{dR}  \bar D_i^L{d_j}- m_{ij}^{\ell R}  \bar E_i^L{e_j} + h.c.\nonumber\,.
\end{eqnarray}
Here the superscripts $L$ and $R$ denote the chirality of the vector-like fields. Note that $Q_i$, $L_i$, $U_i$, $D_i$ and $E_i$, are embedded into complete representations under the PS gauge group, enforcing equality of the respective mass terms. Without loss of generality, we can work in a basis where $M^L_{ij}$ and $M^R_{ij}$ are diagonal in flavour space, and to a good approximation, the masses of the composite states are universal:
\begin{equation}
M_{ij}^L\approx   M_{ij}^R  \approx M {\delta _{ij}}\,.
\end{equation}
For the terms mixing vector-like fermions with the SM ones we assume for simplicity the absence of mixing with the right-handed SM $SU(2)$ singlets, i.e. $m_{ij}^{fR} =  0$. In RS models without brane-kinetic terms $m_{ij}^{fL}$ and $M_{ij}^{L}$ are diagonal in the same basis. Assuming that the brane-kinetic terms are also diagonal in that basis, one can write 
\begin{equation}
m^{fL} =  \left( {\begin{array}{*{20}{c}}
M_1^f&0&0\\
0&{M_2^f }&0\\
0&0&{M_3^f }
\end{array}} \right)\,.
\end{equation}
In the following, we will assume $M_1$ to be zero or negligibly small \footnote{This is motivated by the fact that in order to explain the anomalies no couplings to electrons are required and that once couplings to electrons are present, there arise stringent bounds from $\mu\to e\gamma$ and $\mu\to 3e$. }. Therefore, the first generation is purely elementary while the second and third generation of left-handed SM quarks and leptons are partially composite. Since all terms are flavour diagonal the problem reduces to diagonalizing several $2\times 2$ matrices mixing SM with vector-like fermions. For the quarks and leptons we achieve this by the transformation
\begin{equation}
\left( {\begin{array}{*{20}{c}}
{{f^i_L}}\\
{{F^1_L}}
\end{array}} \right) \to \left( {\begin{array}{*{20}{c}}
{{c_i^f}}&{ - {s_i^f}}\\
{{s_i^f}}&{{c_i^f}}
\end{array}} \right)\left( {\begin{array}{*{20}{c}}
{{f_L^i}}\\
{{F_L^i}}
\end{array}} \right)
\end{equation}
with $i=2,3$, $f=q,\ell$, $F=Q,L$, and $s_i^{f} = \sin  {{\alpha_i^f}} $ with
 ${\alpha_i^f}=\arctan\left({M_i^f}/{M}\right)$. In order to maintain perturbativity of the fundamental Yukawa couplings on the UV-brane, we restrict all mixing angles to $s_i^{f} \le \sqrt{3}/2$.

After EW symmetry breaking additional mass terms for the fermions, originating from the Yukawa couplings, arise. Now, the $3\times 3$ sub-block of the light fermions is in general not diagonal in the same basis as $M^{fL}_{ij}$ and $m^{fL}_{ij}$.  In order to avoid tree-level flavour changing neutral currents (FCNCs) in the down quark sector, we assume the down Yukawa coupling to be aligned with $m^{qL}$ and $M^{qL}$, i.e. diagonal in the same basis. In the left-handed up-quark sector, tree-level FCNCs are then unavoidable, but are determined and suppressed by the small off-diagonal elements of the Cabibbo-Kobayashi-Maskawa (CKM) matrix.

%\subsection{Couplings to gauge bosons}

With these assumptions, the couplings of KK gauge bosons to fermions are given by
\begin{equation}
L_{ff}^V = i{\bar f_i}{\gamma _\mu }\left( {g_L^{V*}\Gamma _{{f_i}{f_j}}^{VL}{P_L} + g_R^{V*}\Gamma _{{f_i}{f_j}}^{VR}{P_R}} \right){f_j}{V^\mu }\,.
\end{equation}
Here, $V=g,\;W^{\pm},\;W^{3},\;{\rm B\!-\!L},\;{\rm LQ}$ with the coupling
 $g^{V*}_{L,R} = \theta g^V_{\rm SM}$ being enhanced by the RS volume $\theta=\sqrt{k\pi R} \sim 6$ with respect to the elementary gauge coupling of the gauge boson $V_{\rm SM}$ . For the LQ and the B-L gauge boson $g^{\rm LQ*}_L =\theta\frac{g_s}{\sqrt{2}}$ and $g^{\rm B-L*}_L =\theta\frac{\sqrt{3}}{2\sqrt{2}}{g_s}$, respectively. The couplings of the KK modes of $SU(2)_R$ to the SM fermions are small.

The relevant matrices in flavour space $\Gamma^{L,V}_{ij}$ read:
\begin{eqnarray}
\Gamma _{{d_i}{d_j}}^{{V^0},L} &=& {\left( {\begin{array}{*{20}{c}}
		0&0&0\\
		0&{s_2^{q2}}&0\\
		0&0&{s_3^{q2}}
		\end{array}} \right)_{ij}} %+\frac{\delta_{ij}}{k\pi R}\,,
\label{eq:7}
	\\
\Gamma _{{u_i}{u_j}}^{{V^0},L} &=& {V_{ik}^{\rm CKM}}{\left( {\begin{array}{*{20}{c}}
		0&0&0\\
		0&{s_2^{q2}}&0\\
		0&0&{s_3^{q2}}
		\end{array}} \right)_{kl}}V_{jl}^{^{\rm CKM}*}%+\frac{\delta_{ij}}{k\pi R}\,,
	\\
\Gamma _{{d_i}{\ell _j}}^{LQ,L} &=& \left( \begin{array}{*{20}{c}}
		0&0&0\\
		0&{s_2^{q}s_2^{\ell}{c_\ell }}& { {s_2^{q}s_2^{\ell}{s_\ell }}}\\
		0&{ - {s_3^{q}s_3^{\ell}{s_\ell }}} &{s_3^{q}s_3^{\ell}{c_\ell }}
		\end{array} \right)_{ij}\,,
\\
\Gamma _{{u_i}{d_j}}^{W,L} &=& {V_{ik}^{\rm CKM}}{\left( {\begin{array}{*{20}{c}}
		0&0&0\\
		0&{s_2^{q2}}&0\\
		0&0&{s_3^{q2}}
		\end{array}} \right)_{kj}}\,,
	%+ \frac{V_{ij}^{\rm CKM}}{k\pi R}
	\\
\Gamma _{{\ell _i}{\ell _j}}^{V,L} &=& {\left( {\begin{array}{*{20}{c}}
		0&0&0\\
		0&{s_2^{\ell 2}c_\ell ^2 + s_3^{\ell 2}s_\ell ^2}&{\left( {s_2^{\ell 2} - s_3^{\ell 2}} \right)} s_\ell ^{}c_\ell ^{} \\
		0&{{\left( {s_2^{\ell 2} - s_3^{\ell 2}} \right)}s_\ell ^{}c_\ell ^{}}&{s_3^{\ell 2}c_\ell ^2 + s_2^{\ell 2}s_\ell ^2}
		\end{array}} \right)_{ij}}\label{eq:11}
	%+\frac{\delta_{ij}}{k\pi R}\\
\end{eqnarray}
Here we neglected flavour mixing with the first generation and dropped the flavour-universal $\theta^2$-suppressed terms~\cite{Contino:2006nn}. In this limit $\Gamma^{R,V}_{ij}=0$. $V^{\rm CKM}$ denotes the CKM matrix, $V^0$ stands for the electrically neutral gauge bosons and $s_\ell$ parametrizes the misalignment in flaovur space between $m^{\ell L}$, $M^{\ell L}$ and the lepton Yukawa coupling in the 2-3 sector.  {Using Eqs.\ \eqref{eq:7}-\eqref{eq:11} one can see that under our assumptions no effects in $K_L\to\mu e$ or $K\to\pi\mu e$ are generated.}

\section{Observables}

\subsection{\boldmath $R(D)$ and $R(D^*)$}

We define the effective Hamiltonian for $b\to c\ell\nu$ transitions as
\begin{equation}
{H_{{\rm{eff}}}^{\ell_f\nu_i}} = \frac{{4{G_F}}}{{\sqrt 2 }}{V_{cb}}C_L^{fi}\left[ {\bar c{\gamma ^\mu }{P_L}b} \right]\left[ {{{\bar \ell }_f}{\gamma _\mu }{P_L}{\nu _i}} \right]\,,
\end{equation}
where in the SM $C_L^{fi}=\delta_{fi}$ and the contribution of our model is given by
\begin{equation}
{C_L^{fi}} = \frac{{\sqrt 2 }}{{4{G_F}{V_{cb}}}}\frac{{\kappa _{3f}^*V_{2k}^{}{\kappa _{ki}}}}{{{M^2}}}+\theta \Gamma^{W,L}_{\ell_f\ell_i} \frac{m_W^2}{M^2}\,.
\end{equation}
Here the first term originates from the LQ, with $\kappa _{ij}=\theta \frac{g_s}{\sqrt 2} \Gamma _{{d_i}{\ell_j}}^{{\rm LQ},L} $, while the second term is due to the KK mode of the $W^\pm$.  Thus we find
\begin{equation}
R(X)/R(X)_{\rm SM}=|1+C_L^{33}|^2+\sum_{i=1}^2|C_L^{3i}|^2\,,
\end{equation}
with $X=D,D^*,J/\Psi$.

This has to be compared to the experimental measurements of $R(D)$, $R(D^*)$ and $R(J/\Psi)$~\cite{Aaij:2017tyk}. A global fit assuming new physics (NP) in $C_L$ only gives~\cite{Alok:2017qsi}
\begin{equation}
C_L^{\rm NP}=0.131\pm 0.033\,.
\end{equation}

\subsection{\boldmath $b\to s\ell^+\ell^-$ transitions}

Using the effective Hamiltonian
\begin{align}
H_{\rm eff}^{\ell_f\ell_i}=- \dfrac{ 4 G_F }{\sqrt 2}V_{tb}V_{ts}^{*} \sum\limits_{a = 9,10} C_a^{fi} O_a^{fi}\,
,\nn {O_{9(10)}^{fi}} =\dfrac{\alpha }{4\pi}[\bar s{\gamma ^\mu } P_L b]\,[\bar\ell_f{\gamma _\mu }(\gamma^5)\ell_i] \,,
\label{eq:effHam}
\end{align}
we have
\begin{equation}
C_9^{fi} =  - C_{10}^{fi} = \frac{{ - \sqrt 2 }}{{2{G_F}{V_{tb}}V_{ts}^*}}\frac{\pi }{\alpha }\frac{{\kappa _{2i}^{}\kappa _{3f}^*}}{{{M^2}}}\,.
\end{equation}
The allowed $2\,\sigma$ range is given by~\cite{Capdevila:2017bsm}
\begin{equation}
	-0.37 \geq  C_{9}^{22}=-C_{10}^{22} \geq -0.88\,.
\end{equation}

Concerning lepton flavour violating $B$ decays, we use the  results of Ref.~\cite{Crivellin:2015era} for the analysis of $B\to K^{(*)}\tau\mu$. The only experimental limit for $\mu\tau$ final states is~\cite{Lees:2012zz}
\begin{equation}
	{\rm{Br}}\left[ {B \to K\tau \mu } \right]_{\rm EXP}\leq4.8\times 10^{-5}\,,
\end{equation}
at $90\%$ confidence level, and the corresponding prediction for our case of $C_9=-C_{10}$ reads
\begin{equation}
{\rm{Br}}\left[ {B \to K\tau \mu } \right] = 1.96 \times {10^{ - 8}}\left( {{{\left| {C_9^{23}} \right|}^2} + {{\left| {C_9^{32}} \right|}^2}} \right)\,.
\end{equation}

\subsection{\boldmath $D-\bar D$ mixing}

$D^0-\bar D^0$ mixing receives tree level contributions from the KK modes of the gluon, the $B-L$ gauge boson and the $W^3$. The resulting NP contribution to the effective Hamiltonian
$
H_{\rm eff} =  C_L Q_1 + h.c.
$
is
\begin{equation}
C_L = \frac{\theta^2\left(\frac{3}{4} g_s^2+\frac{1}{2}g_2^2\right)}{4 M^2}\left(V_{cs}V^*_{us}s_2^{q2}+V_{cb}V^*_{ub}s_3^{q2}\right)^2\,,
\end{equation}
with
$
Q_1 = (\bar c \gamma^\mu P_L u)(\bar c\gamma_\mu P_L u)
$. We have for the matrix element
\begin{equation}
M_{12}^{D} = \frac{1}{3} m_D f_D^2 B_1^D(\mu) \eta_D(\mu)  C_L\,,
\end{equation}
with
$
B_1^D(3\,{\rm GeV}) \approx 0.76$~\cite{Riggio:2017hlu}, $\eta_D(3{\rm GeV}) = 0.77$~\cite{Ciuchini:1997bw,Buras:2000if} and $f_D\approx(0.212)\,$GeV~\cite{Bazavov:2014wgs,Carrasco:2014poa}. Using the HFLAV results of CKM 2016~\cite{Amhis:2016xyh}, the imaginary part of the matrix element  should satisfy
\begin{equation}
|{\rm Im}[M^D_{12}]|< 2\times 10^{-16}\,{\rm GeV}\,.
\end{equation}

\subsection{\boldmath$\tau\to3\mu$}

The neutral B-L and $W^{3}$ KK gauge bosons mediate the decay $\tau\to3\mu$. Using the results of \cite{Crivellin:2015era} and neglecting contributions suppressed by $g_Y/\theta$, we find
%\begin{widetext}
%	\[Br\left[ {\tau  \to 3\mu } \right] = \frac{{m_\tau ^5}}{{768{\pi ^3}{\Gamma _\tau }}}\frac{1}{{{M^4}}}\sum\limits_{V = \gamma ,{W^3},{\rm{B - L}}} {\left( {{{\left| {g_L^{V*2}\Gamma _{{\ell _2}{\ell _3}}^{VL}\Gamma _{{\ell _2}{\ell _2}}^{VL}} \right|}^2} + \frac{1}{2}{{\left| {g_L^{V*}g_R^{V*}\Gamma _{{\ell _2}{\ell _3}}^{VL}\Gamma _{{\ell _2}{\ell _2}}^{VR}} \right|}^2}} \right)} \]
%\end{widetext}
\begin{equation}
{\rm Br}\left[ {\tau  \to 3\mu } \right] = \frac{{m_\tau ^5}\tau_\tau}{{768{\pi ^3}}}\frac{1}{{{M^4}}} {{\left| \sum_V^{W^3,{\rm B\!-\!L}}\!\!\!\!{g_L^{V*2}\Gamma _{{\ell _2}{\ell _3}}^{VL}\Gamma _{{\ell _2}{\ell _2}}^{VL}} \right|}^2}\,.
\end{equation}
Here $\tau_\tau$ is the tau lifetime. This result has to be compared to the current experimental bound of $1.2\times 10^{-8}$~\cite{Amhis:2014hma}. 
	
\subsection{\boldmath$B_s-\overline{B}_s$ mixing}

Due to the assumed flavour alignment in the left-handed down-quark sector, our model does not only forbid tree level contributions to $B_s-\overline{B}_s$ mixing, but also makes the one-loop contributions to $B_s-\overline{B}_s$ mixing finite (even in unitary gauge) due to a suppression mechanism similar to the Glashow-Iliopoulos-Maiani (GIM) one.
In addition, flavour violation solely originates from Yukawa couplings. Thus, the effect is very efficiently suppressed by $1/M^4$.
%With $H=C_1\bar s\gamma^\mu P_L b \bar s\gamma_\mu P_L b$ we get
%\begin{equation}
%{C_1} = -\frac{{\kappa _{2s}^{}\kappa _{3s}^*\kappa _{2t}^{}\kappa _{3t}^*}}{{16{\pi ^2}}}\left( %{\frac{{{D_6}}}{{4M_{LQ}^4}} + {D_2}- \frac{{2{D_4}}}{{M_{LQ}^2}}} \right)\,,
%\end{equation}
%using unitary gauge. Here $s,t=1-6$ labels the six charged leptons. The standard loop functions $D_X\equiv D_X\left( {{M_{LQ}},{M_{LQ}},{m_{\ell_s}},{m_{\ell_s}}} \right)$ are given in the appendix. With out coupling structure, the effect is proportional to  and therefore very small.

\subsection{Direct LHC searches}

The most stringent constraints on the KK mass scale stem from direct LHC searches for resonances decaying to $t\bar t$, dijet or $\tau\bar\tau$. $t\bar t$ resonance searches constrain the RS KK gluon mass to be above $3.3\,\text{TeV}$ in the case of bulk fermions and flavor anarchy \cite{Sirunyan:2017uhk}. In our setup, however, the branching ratio into $t\bar t$ final states is significantly smaller than in the flavour-anarchic scenario, so that we can conservatively lower the mass scale of the lightest resonances to $M = 3.0\,\text{TeV}$. 
Due to the reduced branching ratio into $t\bar t$, the dijet final state is relevant in our setup. The most recent CMS constraint on heavy dijet resonances \cite{CMS-PAS-EXO-16-056} is nonetheless still weaker than the aforementioned $t\bar t$ constraint. Both the B-L gauge boson and the $W^3$ KK mode contribute to the $\tau\bar\tau$ final state. Comparing the $Z'$ of the sequential SM, for which \cite{Aaboud:2017sjh} finds $M_{Z'}>2.42\,\text{TeV}$, with our model, we find that the larger branching ratio into the $\tau\bar\tau$ final state is counteracted by a significantly reduced production cross section: first generation quarks do not couple to the B-L gauge boson, and their coupling to the $W^3$ KK mode is suppressed by $1/\theta$. The $t\bar t$ resonance constraint of $\approx3\,$TeV is hence the strongest limit on the KK mass scale $M$.

\section{Phenomenology}
\label{pheno}

\begin{figure*}[t]
	\begin{center}
		\begin{tabular}{cp{7mm}c}
			\includegraphics[width=0.49\textwidth]{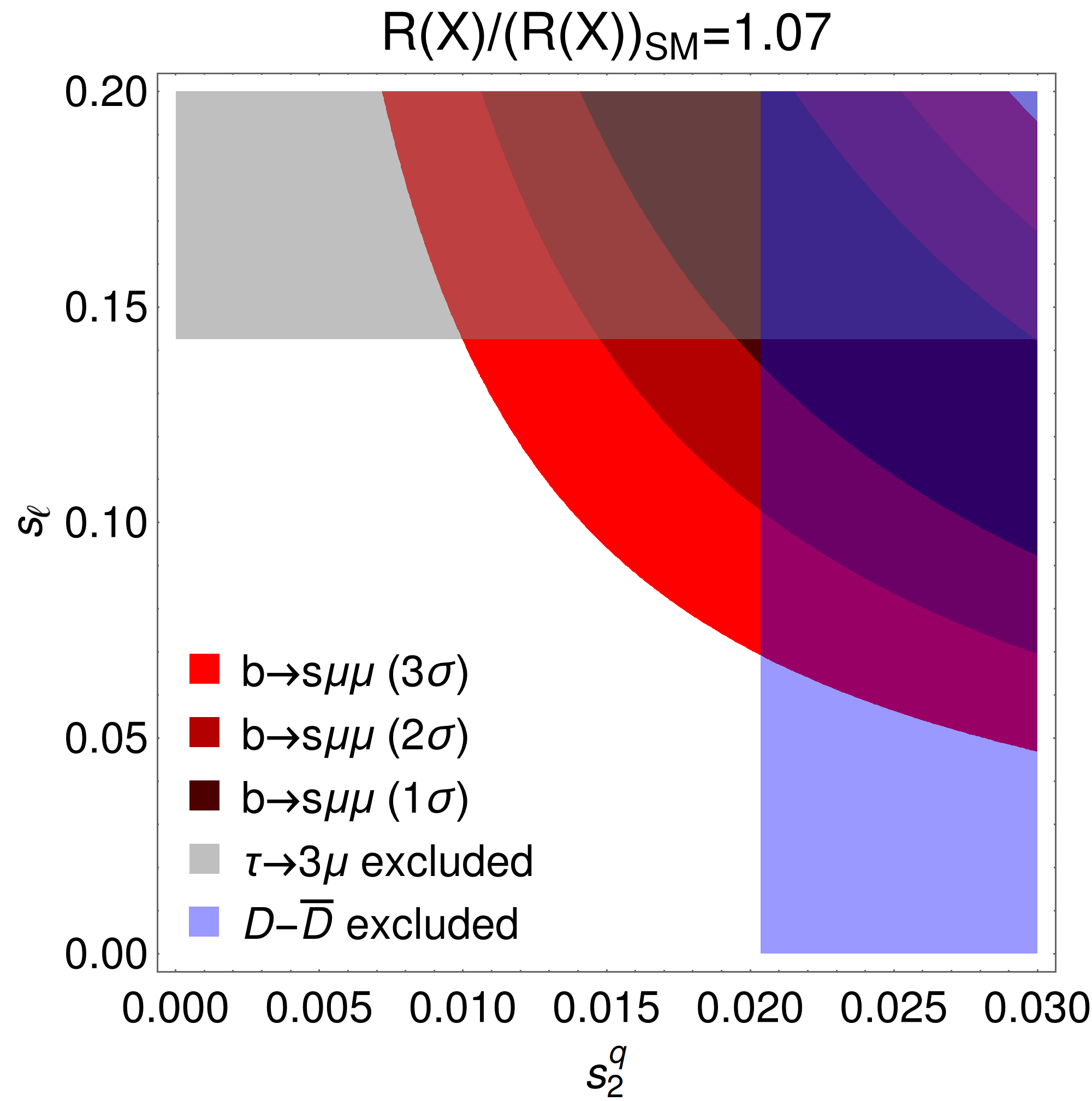}
			\includegraphics[width=0.49\textwidth]{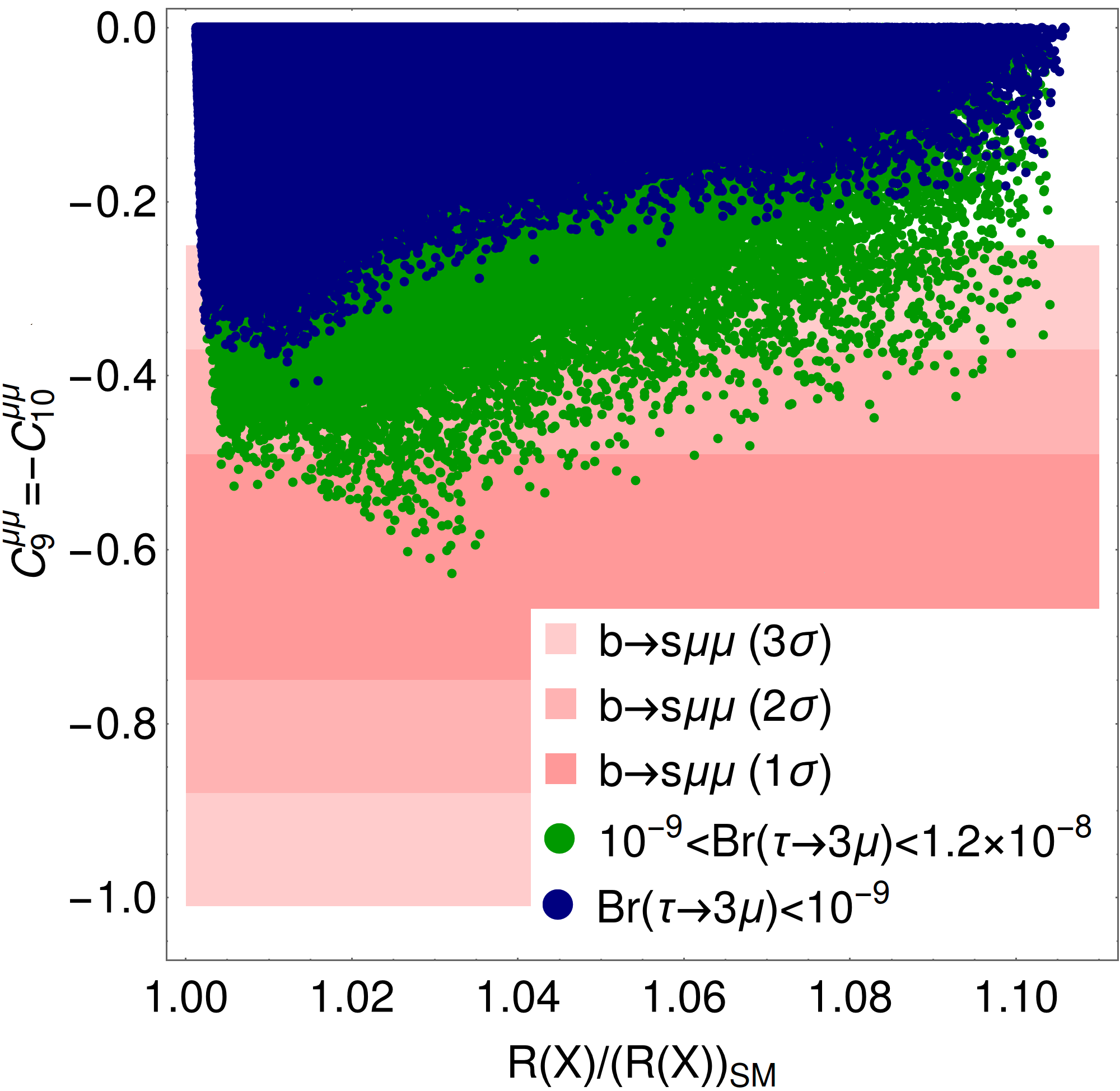}
		\end{tabular}
	\end{center}
	\caption{Left: Allowed regions from $b\to s\mu^+\mu^-$ (red) and the exclusion limits from $D-\bar D$ mixing (blue) and $\tau\to 3\mu$ (gray) for $M=3\,$TeV, $s_{2}^\ell=0.2$, $s_3^\ell=1/\sqrt{2}$ and $s_3^q=\sqrt{3}/2$. With these values, $R(X)/R(X)_{\rm SM}\approx 1.07$ (with $X=D,D^*,J/\Psi$). We see that $b\to s\mu^+\mu^-$ can be explained at the $1\,\sigma$ level without violating the bounds from other observables. 
Right: Correlations between $R(X)/R(X)_{\rm SM}$ and $b\to s\mu^+\mu^-$ for $M=3\,$TeV. Here we scanned over $0.3<s_3^q<\sqrt 3/2$, $0<s_2^q<0.2$, $0.3<s_3^\ell<\sqrt 3/2$, $0<s_2^\ell<0.2$ and $0<s_\ell<0.3$. Only the parameter points consistent with $D-\bar D$ mixing are shown. As we can see, the predicted branching ratio for $\tau\to 3\mu$ is large and very well within the reach of Belle II. }   
	\label{Fig1}
\end{figure*}

Since we aim at getting a large effect in $b\to c\tau\nu$ transitions, a large compositeness of the third generation is required. In addition, $M$ should not be too large and we therefore use a mass of 3$\,$TeV. In order to get a sizeable effect in {$b\to s\mu^+\mu^-$}, while not violating the upper limit on the $\tau\to 3\mu$ branching ratio, moderate values of $s_{2}^\ell$ are preferred. In the left plot of Fig.~\ref{Fig1} we therefore show the allowed regions in the $s_2^q$-$s_\ell$ plane for $s_{2}^\ell=0.2$, $s_3^\ell=1/\sqrt{2}$ and $s_3^q=\sqrt{3}/2$. At this benchmark point $R(X)/R(X)_{\rm SM}\approx 1.07$. {Due to the small coupling to muons (compared to the one to taus) NP effects in $b\to c\mu\nu$ are found below the permille level and therefore consistent with current data~\cite{Jung:2018lfu}.} One can see that $b\to s\mu^+\mu^-$ can be explained at the $1\,\sigma$ level without violating bounds from $D-\bar D$ mixing or $\tau\to 3\mu$.

In the right plot of Fig.~\ref{Fig1} we show the correlations between $R(X)/R(X)_{\rm SM}$ and $b\to s\mu^+\mu^-$ by scanning over $s_3^q$, $s_2^q$, $s_3^\ell$, $s_2^\ell$ and $s_\ell$. Only the {parameter} points consistent with all experimental bounds are shown. We see that in general a large effect in $b\to s\mu^+\mu^-$ limits the size of the possible effect in $R(X)/R(X)_{\rm SM}$ and vice versa. Furthermore, the solution of the $b\to s\mu^+\mu^-$ anomaly in our model predicts a large branching ratio for $\tau\to 3\mu$ within the reach of Belle II.

Due to the constraints from $D^0-\bar D^0$ mixing and $\tau\to3\mu$, we do not {obtain} sizeable effects neither in  $b\to s\tau^+\mu^-$ nor $\tau\to\phi\mu$ \cite{Bhattacharya:2016mcc}, nor in $b\to s\tau^+\tau^-$ transitions as recently examined in Ref.~\cite{Capdevila:2017iqn}. 

\section{Conclusions and outlook}\label{conclusions}

In this article we considered a PS model embedded in the RS space-time in which the symmetry is broken down to the SM one by boundary conditions on the endpoints of the extra dimension. While in previous models based on the PS symmetry the effect in $b\to c\tau\nu$ was only generated by the vector-leptoquark (VLQ), we have as well a $W^\prime$ contribution which enhances in our setup the total NP effect in $b\to c\tau\nu$ processes by roughly 80\%. Still, we find that one cannot fully account for $b\to c\tau\nu$ data due to the stringent constraints from $D-\bar D$ mixing. However, an $O(5\%)$ effect in $R(X)/R(X)_{\rm SM}$ is possible. Furthermore, the model can naturally explain the anomaly in $b\to s\mu^+\mu^-$ transitions including the hints for the violation of lepton flavour universality from $R(K)$ and $R(K^*)$. In addition, our model predicts small effect in $b\to s\tau\mu$  and $b\to s\tau\tau$ transitions, while the effect in $\tau\to3\mu$ is sizable and also the CP violation in the $D-\bar D$ system is close to the current experimental values.

Compared to previous approaches of explaining the flavour anomalies, our model has several advantages. First of all, on the theoretical side, the existence of a massive VLQ and vector-like fermions of the same mass scale follows from the very simple assumption that the PS symmetry is broken on an extra dimension. On the phenomenological side, our model has suppressed couplings of the new particles to light fermions (contrary to Ref.~\cite{Calibbi:2017qbu}) and is therefore quite safe concerning LHC searches. Furthermore, since we have in addition to the VLQ a $W^\prime$ boson which interferes in $b\to c\tau\nu$ processes constructively, we can get an effect which is around 80\% larger compared to the pure VLQ case~\cite{DiLuzio:2017vat,Bordone:2017bld} while still respecting the bounds from $D-\bar D$ mixing.

In our minimal setup we assumed right-handed fermions and the Higgs to be elementary. Giving up these assumptions, one obtains an even richer phenomenology, and also an explanation of the tensions in the anomalous magnetic moment of the muon \cite{ColuccioLeskow:2016dox,Crivellin:2017dsk} and/or in $\varepsilon'/\varepsilon$ \cite{Bobeth:2017ecx} could become possible.

%%%%%%%%%%%%%%%%%%%%%%%%%%%%%%
%\medskip
%%%%%%%%%%%%%%%%%%%%%%%%%%%%%%

{\it Acknowledgments} --- {\small M.B. thanks PSI for the warm hospitality during her visits leading to this publication. The work of A.C. is supported by an Ambizione Grant of the Swiss National Science Foundation (PZ00P2\_154834).}

\bibliography{BIB}

\end{document}